\documentclass[review]{article}
\usepackage{graphicx}
\usepackage[cmex10]{amsmath}
\usepackage{caption}
\usepackage{amssymb}
\begin{document}
\title{Asynchronous Operations on Qubits in Distributed Simulation Environments using QooSim}

\author{J. Joel vanBrandwijk and Abhishek Parakh\\
Nebraska University Center for Information Assurance\\
School of Interdisciplinary Informatics\\
University of Nebraska\\
Omaha, NE 68182}

\maketitle

\begin{abstract}
Operations on a pair of entangled qubits are conventionally presented as the application of the tensor product of operations. The tensor product is linearly extended to act synchronously across the entire entangled system. When simulating an entangled system, the conventional approach is possible and practical if both parts of the entangled system exist within the same physical simulator. However, if we wish to simulate an entangled system across a distributed network, sending half of the entangled pair to another simulator on another computer system, the synchronous approach becomes difficult. In the first part of this paper, we demonstrate a method of simulating operations on entangled states in a distributed environment which is equivalent to the conventional approach. The advantage to our approach is that we can simulate distributed quantum systems on physically distributed hardware. Such a system advances the possibilities of demonstrating distributed quantum algorithms for research, teaching, and learning.

Further, the security of quantum key exchange depends on successfully detecting the presence of an eavesdropper. In most cases this is done by comparing the errors introduced by an eavesdropper with the channel error rate. In other words, the communicating parties must tolerate some errors without losing a significant amount of key information. In the second part of this paper, we characterize the effects of amplitude damping errors on quantum key distribution protocols and explore allowable tolerances. Through simulations we observe that the effect of these errors, in some cases, is highly dependent on where the eavesdropper is located on the channel.

In this paper, we also briefly describe the development of a new quantum simulation library called QooSim.
\end{abstract}

\section{Introduction}
Quantum computing refers to the use of quantum mechanics to construct a circuit or machine which transforms or transmits data \cite{Gershenfeld}. The field of quantum computation holds many exciting promises. Shor's algorithm can break current state of the art cryptography, but many quantum algorithms have been developed to provide unbreakable encryption based on the laws of physics \cite{Shor} \cite{bb84}.

The subject of quantum computing, however, remains difficult for learners and new researchers to approach. At it's base, quantum computing relies on an understanding of quantum mechanical principles such as superposition, no-cloning, measurement, and entanglement. However, these topics run counter to classical knowledge about what a computing machine should be and how it should behave. Where classical models describe information as a series of discrete bits (zeros and ones), quantum computing instead has a single quantum bit, or qubit, represented by probabilities of zero or one - while existing as both until measured \cite{desurvire}.

Adding to the conceptual difficulties encumbering learning and research in the field of quantum computing are technical and logistical problems. To begin with, commercial implementations of quantum computing hardware still only performs on par with standard, consumer grade classical hardware \cite{Dwave}. As quantum computing hardware is difficult to implement, it is also \emph{rarely} implemented outside of specialized laboratories. Access to a quantum computer is not within the reach of most researchers or students.

As a result simulators capable of simulating quantum computing principles are useful for educational purposes. It is important to realize that these ``quantum simulators" are not emulators of the underlying physical principles of quantum computers, but rather mathematical models of how quantum bits are expected to behave and interact. Even so, quantum simulators are invaluable tools in teaching, learning, and researching quantum principles and algorithms.

In this paper, we describe the development of a new quantum simulator library called QooSim. The simulator is currently focused on simulating quantum key distribution (QKD) protocols along with error correction involved in QKD. In order to do so we developed several of the basic technique needed for the simulation of a general quantum computing environment such as superposition, measurement, application of quantum gates, quantum error correction, etc.

Besides describing the general structure of the QooSim, we will focus specifically on the technique that we developed for simulating quantum entanglement. Since quantum key distribution assumes geographically distributed parties, we found that simulating quantum entanglement when individual qubits of the entangled system may exist at different locations over a network is particularly challenging. The technique that we present in this paper to solve this problem is one of the main contributions of the paper.

In section \ref{sec:quantumEntanglement}, we discuss the notion of entanglement in quantum computing. Section \ref{sec:previousWork} introduces some of the available quantum simulators in order to explain the missing features which motivated our own effort. Section \ref{sec:linearExtension} describes the conventional approach to mathematically modeling entanglement, and section \ref{sec:asyncOperations} describes our approach when implementing the simulator. These methods are presented side by side as verification of equivalence. Section \ref{sec:QooSimOverview} provides technical information about the data structures and algorithms used to construct our simulation library. Section \ref{sec:experimentalResultBB84} provides experimental results of our simulator, which demonstrates the equivalence between our approach in the simulator and the theoretical approach. In section \ref{sec:simulatingAmplitudeDamping} we introduce the amplitude damping error model for quantum channels. In section \ref{sec:analysis} we describe the various test cases executed, and discuss the results of each. Section \ref{sec:implications} describes the practical implications of these results in light of recent experimental observations. The paper concludes in section \ref{sec:conclusion}.

\section{Quantum Entanglement}\label{sec:quantumEntanglement}

Any pair of qubits, $P$ and $Q$, can be represented in the general form
\begin{equation}\label{eq:R}
|R\rangle = \alpha|00\rangle + \beta|01\rangle + \gamma|10\rangle + \delta|11\rangle
\end{equation}
where $\alpha, \beta, \gamma,$ and $\delta$ are the probability amplitudes of their associated states \cite{desurvire}. The probability of a state occurring on measurement is given by the square of the absolute value of the probability amplitude. Thus, the only condition on the general two qubit state is that the sum of the squares of the probability amplitudes equal one.

An entangled state of two qubits has the special condition that there exists no tensor product of pure single qubit states capable of generating the entangled state \cite{desurvire}. To describe entanglement in another way, there is no way to factor an entangled state into composite pure states. In terms of the equation \ref{eq:R}, at least one of the state amplitudes must be zero, but no two amplitudes may be zero in such a way that the value of one of the pair of entangled qubits is determined. This rule can be stated as follows, let
\[ A = \left( \begin{array}{cc}
\alpha & \beta \\
\gamma & \delta \end{array} \right)\\
\]
Then if $A$ satisfies the condition that the sum of none of the rows or columns is zero, $A$ represents the probability amplitudes of a two qubit entangled state.

A special subset of entangled states are the Einstein, Podolsky, Rosen (EPR) or Bell states \cite{desurvire}. The Bell states represent the four maximally entangled states of two qubits. These four states are:

\[1/\sqrt{2} (|00\rangle \pm |11\rangle) \]
\[1/\sqrt{2} (|01\rangle \pm |10\rangle) \]

\begin{figure}
\begin{center}
\includegraphics[width=\linewidth]{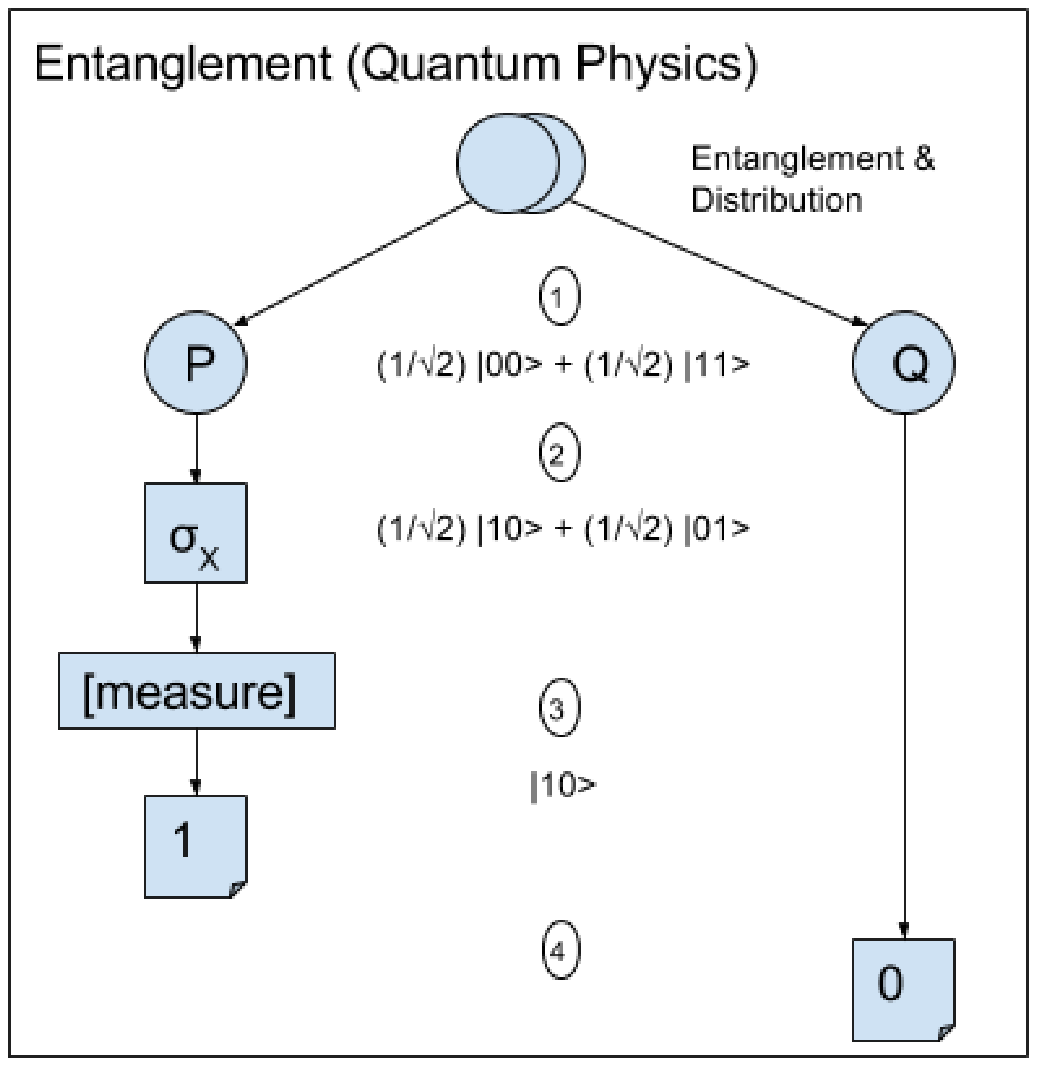}
\end{center}
\captionsetup{font={small,it},singlelinecheck=off}
\caption[goo]{
\begin{enumerate}
\item A pair of qubits is entangled in such a way that if one qubit is measured to be $|0\rangle$, the state of the system collapses to $|00\rangle$. Likewise for the measurement of $|1\rangle$, the collapse will be $|11\rangle$. There is an equal chance of either measurement. The system is in the state $(1/\sqrt{2})|00\rangle + (1/\sqrt{2})|11\rangle$
\item A sigma-x gate is applied to $P$, changing the state of the system so that if $P$ is measured to be $|0\rangle$, the state of the system collapses to $|01\rangle$. If $P$ is measured to be $|1\rangle$, the state of the system collapses to $|10\rangle$. Likewise for $Q$. The system is in the state $(1/\sqrt{2})|10\rangle + (1/\sqrt{2})|01\rangle$
\item A measurement is performed on $P$, resulting in $|1\rangle$. The system collapses to $|10\rangle$
\item As a result, if measured in the same basis, $Q$ will be measured as 0.
\end{enumerate}
}
\label{figPhysEnt}
\end{figure}

That is, when one of the pair is measured, the value of the other member of the pair, if measured in the same basis, is completely determined \cite{Veritasium}. An illustration of this phenomena is shown in figure \ref{figPhysEnt}. This effect occurs instantaneously no matter the physical distance between the two qubits.

The high cost and limited availability of quantum computing hardware presents a significant barrier to experimentation with these phenomena. We set out to create a tool reasonably capable of simulating these effects on classical hardware. The challenge of simulating ``spooky action" led us to explore new methods of operating on entangled systems.

Nonetheless, entanglement has been experimentally verified time and again, most recently by astronauts on the International Space Station \cite{ESA}. The seemingly odd ``spooky action" phenomenon of entanglement turns out not only to be true, but to play a central role in the effort to make practical quantum networks and protocols as well. As with all signals, quantum signals degrade as they travel through a medium. Entanglement has been shown to have the potential to correct errors in which the bit values or phases are flipped \cite{shorCode} \cite{shorCodeImpl} \cite{Gottesman}. In our current research, we explore the effects of a similar entanglement based error correction mechanism on partial rotational errors \cite{correctRotation}. A ``dual-rail" system has been proposed using entangled qubits to solve amplitude damping errors \cite{dualRail}. Entanglement has been theorized as the mechanism behind quantum ``repeaters", which promise to solve the problem of amplifying a signal which cannot be read without being destroyed \cite{repeater} \cite{repeaterImpl}.

Quantum entanglement can be used to improve non-entangled quantum encryption schemes such as BB84 \cite{Parakh, Ekert}, or to develop new encryption schemes, such as Ping-Pong algorithm. Superdense coding, which allows for more efficient data transfer and quantum reexportation, are both possible because of quantum entanglement \cite{Branciard, Bennett}.

\section{Previous Work}\label{sec:previousWork}
In this section, we provide a survey of existing quantum simulators. We do so in an effort to show that while there are certainly very well constructed quantum computing libraries available, two features are rarely implemented: distributed simulation across a quantum channel and behavior of entangled qubits. We further demonstrate that the union of these features, simulation of the behavior of entangled qubits across distributed systems, has not, to our knowledge, been addressed. Such a feature would enable the simulation of entanglement based quantum key distribution protocols on classical hardware. We believe this to be an important tool for validating security and efficiency claims about such protocols.

Libquantum is one of the most widely used and cited \cite{libquantumBib} quantum computing libraries. Indeed, the preliminary work on our quantum library was to port libquantum to C++. However, as we will later show, our work has since evolved and bears little resemblance to structure of the original.

Libquantum is written in C, and received its last major update in 2013. The library is structured around quantum registers. Quantum registers contain a set of nodes. Each node corresponds in turn to a possible value for the register and the probability of that value occurring. The authors chose this design, presumably, for greater efficiency with quantum algorithms such as Shor's and Grover's. Example implementations of both algorithms are included in the library. However, a consequence of this structure is that storage of quantum registers consumes system memory at an exponential rate. Indeed, we struggled to use a quantum register longer than 22 qubits on our development machines.

Libquantum contains a number of built-in gates and operations for manipulating quantum registers, including Hadamard, sigma and rotation gates. The library also supports controlled operations on pairs of qubits, such as swap and CNOT gates and on 3-tuples of qubits, such as the Toffoli gate. It does not, however, have facilities for sending these registers across a network or even for serialization of these structures. Together with a limited register size, it is difficult to simulate entanglement based quantum key distribution algorithms using libquantum across a network.

Our initial efforts to extend libquantum involved lexical serialization and re-assembly of data structures into raw TCP data streams. Operations were then performed by sending specialized code-word tokens, again over raw TCP data streams. We had in essence begun to design an application layer protocol for libquantum. However, this effort quickly proved cumbersome and difficult to maintain and extend. The desire to avoid creation of a network based protocol exclusively to support our simulator became one of the design goals for our library. After all, serialization is a problem which has been solved many times over and besides not within the scope of our problem.

Another simulated called Q++ is a somewhat later addition to the family of quantum simulators. As the name implies, the library is written in C++. The last update to this library was in 2013. The library is designed using a series of templates, allowing the user to construct one or more quantum ``simulator" objects, each with one or more quantum ``register" objects associated. The internal representation of quantum information is the same as for libquantum; a list of possible states is stored in correlation with the probability of each state occurring. Q++ has a feature for defining quantum operations, or QOPs, which are in turn a construction of one or more quantum gates, similar to a quantum circuit.

The authors were able to compile the library, but were limited in our ability to test this library. We suspect that Q++ would have the same memory consumption issues encountered by libquantum. Q++ does not have facilities for network transmission or serialization of quantum information. While the design of Q++ shows some exciting promise with simulator objects, we believe it would be difficult to simulate entanglement based quantum key distribution protocols using Q++. We did learn from this approach that creating a control object to manage registers would be a desirable approach. We incorporated this concept into our design through the ``System" object.

Quantum simulation is not limited to strongly typed, compiled languages. The Quantum::Entanglement module aims to ``port some of the functionality of the universe into Perl" \cite{PerlQuantumEntanglement}. The module is unique in that it allows any set of states to be represented probabilistically, including floats, integers, strings, and objects. Any numerical probability can be assigned to these states. The library will self-normalize the probabilities. Quantum::Entanglement not only allows two quantum probabilistic values to be related, but allows for the ``entanglement" of classical values onto an entangled state. In doing so, users are able to change the meaning of a quantum state on the fly, and create complex classical-quantum hybrid algorithms.

Being written in Perl, native serialization libraries would certainly be capable of encoding Quantum::Entanglement variables for transmission. Indeed, native protocols would be capable of transmitting these serialized variables. However, once transmitted, the quantum variables are decoupled. The library does not include a mechanism for simulating entangled states across distributed systems.

While Quantum::Entanglement has an interesting set of features and met our requirement for serialization of data structures, we chose not to pursue it as a basis for our simulator because of the weak variable binding and lack of native support for quantum gates.

We examined three other quantum simulators during our research. QCL, jQuantum, and pyQu each take a different approach to the task of quantum simulation. It should be said that all of the tools we examined are commendable for their innovation and highly recommended to anyone wanting to explore more with quantum computing algorithms and concepts. We encourage readers to test out each of these libraries for themselves. Our research has a very specific goal and intention; that no existing tool could meet this goal should not diminish these works in the least. Nor do we claim to have superseded any existing simulator. We have simply solved a different aspect of the simulation problem.

\section{Linear Extension Operations}\label{sec:linearExtension}
Any operation on one qubit in an entangled pair state necessarily has an effect on the other qubit (figure \ref{figPhysEnt}). For unentangled qubits, gates operations are the product of the matrix representation of the gate ($G$) with the matrix representation of the qubit \cite{Vazirani, Moore}. $G$ is a unitary matrix, with coefficients $a$ and $b$ such that $|a|^2 + |b|^2 = 1$. To operate on the generic pair state, $R$, the tensor product of $G$ and the identity matrix $I$ are used to map a $2 \times 2$ unitary matrix onto a $4 \times 4$ matrix \cite{Vazirani, Moore}.

Given
\[
R = \alpha|00\rangle + \beta|01\rangle + \gamma|10\rangle + \delta|11\rangle\\
\]
and
\begin{equation}
\label{eq:G}
G = e^{\varphi} \left( \begin{array}{cc}
a & b \\
-b^* & a^* \end{array} \right)\\
\end{equation}

\[
(I \otimes G) R = e^{\varphi} \left( \begin{array}{cccc}
a & b & 0 & 0\\
-b^* & a^* & 0 & 0 \\
0 & 0 & a & b \\
0 & 0 & -b^* & a^* \end{array} \right)
\left( \begin{array}{c}
\alpha\\
\beta\\
\gamma\\
\delta \end{array} \right)
\]

\[
(I \otimes G) R
= e^{i \varphi} \left( \begin{array}{c}
\alpha a + \beta b \\
\alpha (-b^*) + \beta (a^*) \\
\gamma a + \delta b \\
\gamma (-b^*) + \delta (a^*) \end{array} \right)
\]

\begin{equation}
\label{eq:sync}
\begin{split}
(I \otimes G) R &=
e^{i \varphi}(\alpha a + \beta b) |00\rangle \\
&+ (\alpha (-b^*) + \beta (a^*)) |01\rangle \\
&+ (\gamma a + \delta b) |10\rangle \\
&+ (\gamma (-b^*) + \delta (a^*)) |11\rangle
\end{split}
\end{equation}

\section{Description of Asynchronous Operations}\label{sec:asyncOperations}
When simulating quantum systems and algorithms, it is necessary to use classical hardware and methods. This includes simulation of entangled qubits distributed across multiple computer systems over a classical network. It is in many cases necessary to apply quantum gates to these entangled systems after they have been dispersed across physical systems.

The conventional method of applying the tensor product of the gate and the identity matrix becomes challenging when considering these conditions. Possible solutions would include maintaining both qubits from a pair on both systems, applying the operation in parallel on each system. Alternatively, a broker service could be used to manage information until measurement. However, we will show that there exists an equivalent asynchronous operation on a single qubit which can be used to simulate a series of synchronous operations on the system.

\begin{figure}
\begin{center}
\includegraphics[width=\linewidth]{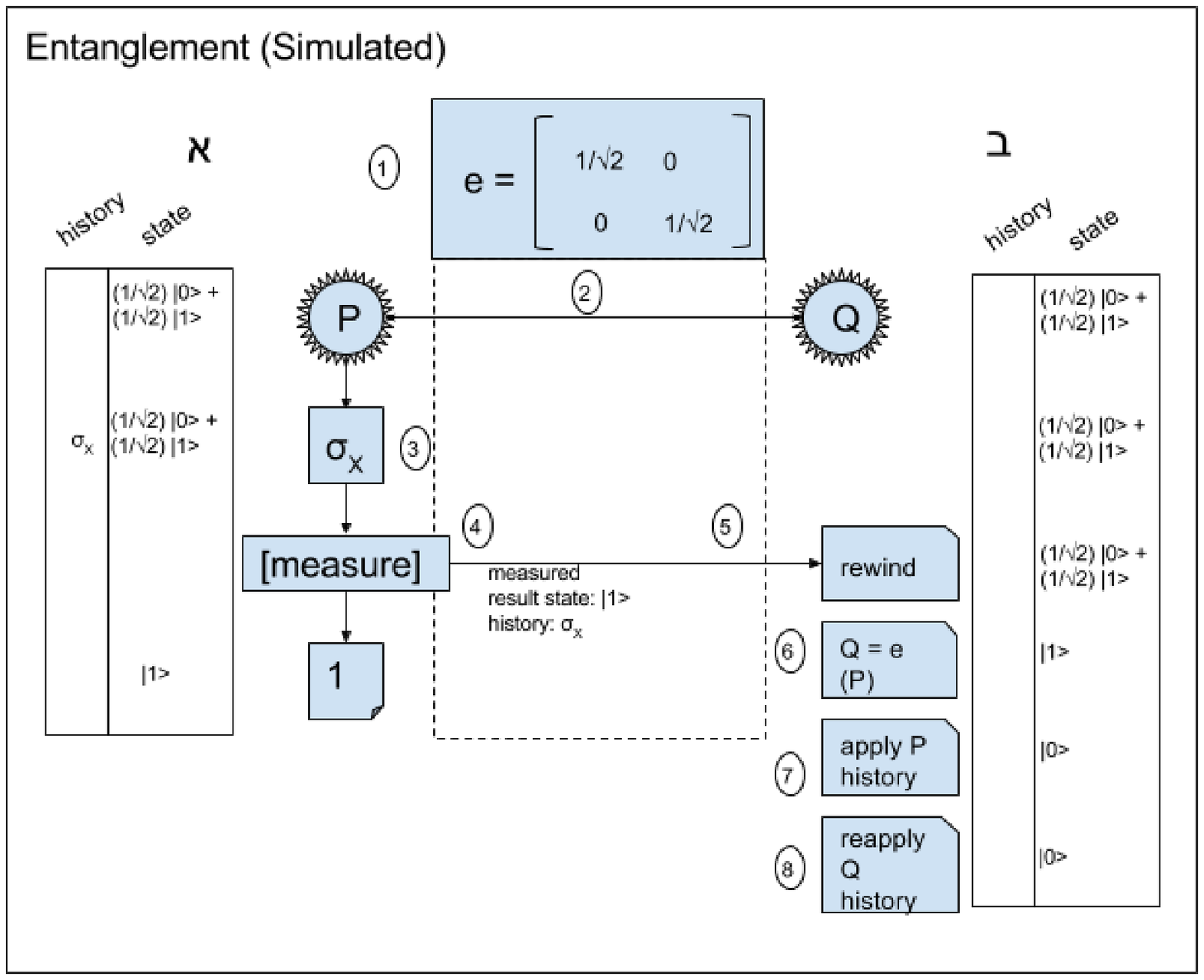}
\end{center}
\captionsetup{font={small,it},singlelinecheck=off}
\caption[goo]{
\begin{enumerate}
\item The EPR matrix in the entanglement object is set to $\alpha = 1/\sqrt{2}, \beta = 0, \gamma = 0, \delta = 1/\sqrt{2}$, which corresponds to the physical system state $(1/\sqrt{2})|00\rangle + (1/\sqrt{2})|11\rangle$
\item $P$ and $Q$ are initialized with their states each at $(1/\sqrt{2})|0\rangle + (1/\sqrt{2})|1\rangle$ and associated to their respective halves of the entanglement object.
\item A sigma-X gate is applied to $P$. $P$ processes the gate operation and stores the matrix corresponding to the gate in its history vector. Note that the gate operation has no discernable affect on the measurement outcome for the individual qubit represented by $P$.
\item A measurement operation is applied to $P$. $P$ measures the qubit to be one, and then notifies the entanglement object of its history and the measurement results.
\item The entanglement object notifies $Q$ that a measurement operation has taken place on $P$. Included in this notification are $P$'s result and history. $Q$ ``rewinds" any operation which has been performed on itself by applying the inverse matrix operation.
\item $Q$ queries the entanglement object to find that in their original states, if $P$ was measured as $|1\rangle$, $Q$ would have also been measured $|1\rangle$. $Q$ sets its state accordingly to $(0)|0\rangle + (1)|1\rangle$.
\item $Q$ now applies $P$'s history, applying the sigma-X gate, which sets its state to $(1)|0\rangle + (0)|1\rangle$.
\item $Q$ would now re-apply any operations which had taken place on it prior to notification from the entanglement object. No operations took place in this scenario. Note that if $Q$ were now measured, the result would be $|0\rangle$ with 1.0 probability.
\end{enumerate}
}
\label{figSimEnt}
\end{figure}

We start by describing the system for simulation, shown in figure \ref{figSimEnt}. At the moment a pair of qubits $P$ and $Q$ are entangled, each system $\aleph, \beth$, respectively must begin maintaining a history of its actions. Additionally, the simulator must maintain a method for each qubit to communicate with the other, such as a network address. Conventional, single qubit gates may be applied without limitation to each qubit in the pair individually for the duration of the entanglement. When entanglement is broken, such as by measurement, the simulators must reconcile the state of their half of the entangled pair.

If the qubit $P$ is measured, $\aleph$ then should alert $\beth$ of the measurement, and inform $\beth$ of all operations performed on $P$ since entanglement but before measurement. Conceptually, $\beth$ should ``rewind" the operation history of $Q$, restoring the initial entangled state. $\beth$ will then update the amplitudes of $Q$ to reflect the outcome of the measurement, and then replay both the operational history of $P$ and $Q$ against $Q$. The result will provide the state of $Q$, disentangled, after the measurement of $P$. A measurement of $Q$ will now match observable real world outcomes.

In the following we prove that performing asynchronous operations on a single qubit in the pair is equivalent to performing the same operation as a linear extension on the entangled pair.

Given (\ref{eq:R}) and (\ref{eq:G}), and assuming $R$ to be composed of the qubits $P = p_0 |0\rangle + p_1 |1\rangle$ and $Q = q_0 |0\rangle + q_1 |1\rangle$ \\
\[
\begin{split}
P (G \otimes Q ) &=
e^{i \varphi} (p_0|0\rangle ((aq_0 + bq_1)|0\rangle \\
&+ ((-b^*)q_0 + (a^*)q_1)|1\rangle) \\
&+ p_1|1\rangle ((aq_0 + bq_1)|0\rangle \\
&+ ((-b^*)q_0 + (a^*)q_1)|1\rangle))
\end{split}
\]
\[
\begin{split}
P (G \otimes Q ) &=
e^{i \varphi} ((ap_0q_0 + bp_0q_1)|00\rangle \\
&+ ((-b^*)p_0q_0 + (a^*)p_0q_1)|01\rangle \\
&+ (ap_1q_0 + bp_1q_1)|10\rangle \\
&+ ((-b^*)p_1q_0 + (a^*)p_1q_1)|11\rangle)
\end{split}
\]
And now, since $p_0q_0 = \alpha$, $p_0q_1 = \beta$, $p_1q_0 = \gamma$, and $p_1q_1 = \delta$:
\begin{equation}
\label{eq:async}
\begin{split}
P (G \otimes Q) &=
e^{i \varphi}(\alpha a + \beta b) |00\rangle \\
&+ (\alpha (-b^*) + \beta (a^*)) |01\rangle \\
&+ (\gamma a + \delta b) |10\rangle \\
&+ (\gamma (-b^*) + \delta (a^*)) |11\rangle
\end{split}
\end{equation}

Since (\ref{eq:sync}) is equivalent to (\ref{eq:async}), the operations on a single qubit are equivalent to the linear extension applied to the system.

It is worth noting that any of the paired state amplitudes, $\alpha...\delta$ could be zero valued, thus encompassing the Bell pair states in this definition.

\section{Overview of QooSim}\label{sec:QooSimOverview}
Our motivation behind QooSim was to create an object oriented library for simulating quantum algorithms and protocols, particularly in a teaching/learning environment. Our primary areas of research are quantum key distribution and quantum error correction. It was therefore important to us that the library be able to serialize and un-serialize data structures for transfer across TCP/IP networks, so that teams of students and researchers could work together on separate nodes. We also knew that we wanted to be able to support entanglement based protocols and error correction mechanisms, so support for operations on entangled qubits was a required feature.

We began with the intention of porting the original libquantum to C++ and implementing object oriented design principles to represent and interact with the C based structures and functions. As such, the initial implementation simply encapsulated and compartmentalized the existing features without adding any additional functionality. We eventually found a number of ways to improve on this design, described in this section.

One of the larger changes we made was to replace the state/amplitude quantum register structure with a matrix based representation. As noted above, this structure limited the size of a register to no more than 22 qubits on practical hardware. This is because as the size of the register grows, the amount of memory required to represent each qubit also grows exponentially. For a 22 qubit register, $2^{22}$ integers and $2^{22}$ complex floats were required to be stored in a C array.

By comparison, our implementation requires a constant two complex floats per qubit. We store each qubit as an independent object, with properties corresponding to the probability amplitudes of an individual qubit being measured as zero ($\alpha$) and of the same qubit being measured as one ($\beta$). We traded the processor efficiency of having each possible state's amplitude cached with the memory efficiency of only storing the individual qubit amplitudes. It should be noted that no information is lost by this decision; we are still able to calculate each state possibility on demand.

Our quantum register class implements a composition design pattern as a collection of qubits objects. As expected, this quantum register shares many design features with the classical register from which it is derived. However, there are also methods implemented for the application of quantum gates on the register, for measurement of the register, and for serialization.

Quantum gates are the basic units of operation on qubits. Quantum gates commonly operate on one or two qubits, and are described by $2 \times 2$ or $4 \times 4$ unitary matrices. We define gates using the template pattern. Several common gates are included: general rotational transformation and Pauli, Hadamard, CNOT and Toffoli gates. Each gate has a matrix, defined either at compile time, such as Hadamard, or at runtime, as in the rotational gates. Users of the library can implement new gates by extending the Gate class.

Gate objects are passed to the quantum registers through applyGate method, along with a positional parameter. Gates cannot be directly applied to the qubit objects, similar to a Command pattern implementation. This was an important design step for us, because to support entanglement on distributed systems, we would need to monitor, capture, and respond to quantum operations on distributed pairs of entangled qubits. By placing the register class in a position to broker these gate operations, we were able to meet this requirement.

Measurement collapses the superposition state $\alpha|0\rangle + \beta|1\rangle$ to the states $|0\rangle$ or $|1\rangle$ for the rectilinear basis. Analogous statements are true for other bases. We decided to implement the measurement operation only for the rectilinear basis. For measurement in other bases, qubits can simply be rotated by the user before measurement; that is, for measurement in base $\theta$, rotate by $-\theta$.

The measurement operation is again implemented by calls through the register level; similar to gate operations, we must be able to monitor and respond to measurement of part of an entangled system. The register defers the actual measurement operation to the qubit object. The qubit object generates a random number using the system's random number generator. If the random number is less than $|\alpha|^2$, the measurement operation returns 0. Otherwise, it returns 1.

To incorporate serialization of registers, we turned to Google protocol buffers, one of the leading technologies in this area. By using protocol buffers, we also gained access to Google Remote Procedure Call (RPC) functionality for supporting services and requests on distributed systems. With this pair of technologies, we had found a way to add network serialization and transmission of quantum registers to our simulation environment.

This lead us to add a new high level class to our library: the System class. The System class follows the Singleton pattern, and acts as a registration point for quantum objects. The class allows us to manage a set of Quantum Register objects, which users of the API can also instantiate independently. The System class also has a Google RPC object to enable sending, receiving, and remote operations on register objects. At this stage in the implementation, we had the ability to simulate quantum protocols which did not require entanglement across a distributed network.

\subsection{Simulating Channel Noise} To understand and evaluate quantum network algorithms in a more meaningful way, we wanted to have the ability to simulate the quantum errors. These errors could be introduced passively by signal noise or actively by an attacker siphoning off and measuring qubits. We wanted to be able to understand the impact of both kinds of noise on quantum key distribution protocols, such as BB84 \cite{bb84} and KAK06 \cite{Kak06}, and study quantum error correction techniques such as Shor code \cite{shorCode} and dual-rail qubits \cite{dualRail}.

In reference to Eve the eavesdropper commonly used when describing the security of cryptographic techniques between Alice and Bob, we implemented error simulation as an interface, iEvil, with a method doEvil. We created a default error class, SilentEvil, which has a doEvil method which simply measures qubits. Qubits are measured at a rate defined by the evilness property, which defaults to .05.

Users of the library can define their own error scenarios by implementing the iEvil interface. For example, one could introduce collective rotational error by applying a rotation gate to each position in the register, or amplitude damping by applying the appropriate matrix as a gate object. In the second case, strictly speaking, since amplitude damping matrices are not unitary, they are not gates. However, within the context of the simulator, we can apply these matrices through the same gate application methods.

iEvil objects are added to communications through the System object, and remain in effect until removed. The System will pass all outgoing quantum registers through the doEvil method before transmission.

\subsection{Entanglement} 
\begin{figure}
\begin{center}
\includegraphics[width=\linewidth]{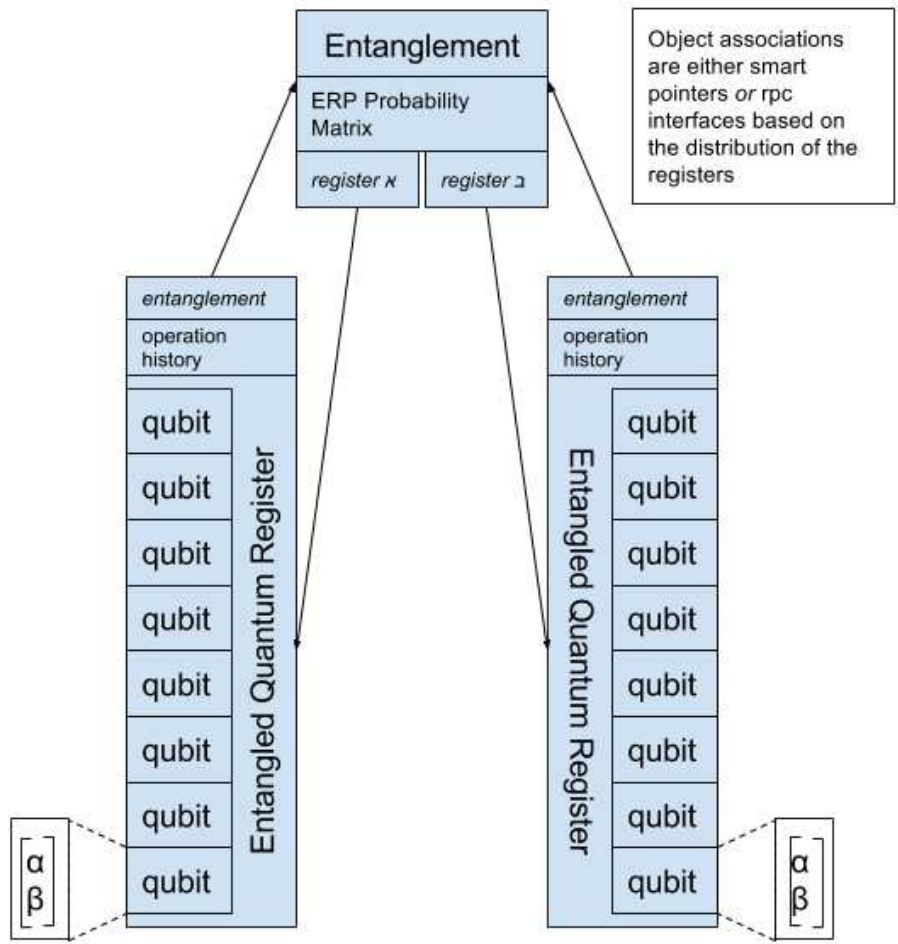}
\end{center}
\captionsetup{font={small,it},singlelinecheck=off}
\caption[goo]{Structure of EntangleRegister and Entanglement classes}
\label{figReg}
\end{figure}

Next, we turned our attention to the question of entanglement. We extended the Register class into the EntangledRegister class, which stores a history of the operations performed on itself. We created a new class, the Entanglement class, to coordinate interactions between two EntangledRegister objects. Each EntangledRegister has a reference to the Entanglement and viceversa. The Entanglement class stores only the original entangled state of each pair of entangled qubits (EPR Probability Matrix); it is not necessary to track changes to the system state. These relationships are shown in figure \ref{figReg}.

By forcing each of a pair of entangled qubits to be in a different EntangledRegister object and implementing well defined interfaces, we neatly staged our problem. Whatever solution we used to simulate entanglement locally, we could extend to our distributed network model through the use of RPC calls. This brings us to the crux of this work: it is necessary to simulate qubits as distinct objects in order to separate them over a network, but entanglement causes a pair of qubits to be treated as a system. How can we simulate entanglement when the entangled qubits do not co-exist within the same simulated object -- or even the same computer system?

By using the distributed interaction model described in section \ref{sec:asyncOperations}, we are able to simulate precisely this scenario. When one register in a pair of entangled registers is sent over the network via protocol buffers, its reference in the Entanglement object is replaced by a stub object. This stub object forwards measurement notifications over the network to the System singleton on the other side of the connection. This notification includes a history of the local quantum operations, which will be applied by the remote system.

An example may be helpful. Suppose Alice prepares an Entanglement to represent the state $1/\sqrt2|00\rangle + 1/\sqrt2|11\rangle$ by creating an entanglement object. Her object contains the matrix $M_e =(1/\sqrt2, 0, 0, 1/\sqrt2)^T$ to describe the entanglement, and two one-qubit registers, one for each qubit in the pair. She keeps the first qubit in this state locally, and sends the second qubit to Bob, who instantiates his own Entanglement object. At this point, Alice and Bob each have their own Entanglement objects with matrix $M_e$, an EntangledRegister, and a stub register representing their respective remote partners.

Alice then applies a Sigma-X gate to her qubit. Her EntanglementRegister's applyGate method catches this action and applies the gate as normal, but also stores the matrix representation of the gate in the opHistory property. For the sake of completeness, suppose that Bob applies a rotational gate, say rotation about $x$-axis by $\pi/3$. Then this operation is caught and stored in Bob's EntangledRegister's opHistory property.

Now, Alice performs a measurement of her qubit. Let us suppose that Alice's measurement returned a 1. Again, her EntanglementRegister catches this measurement and forwards a message to her Entanglement object. Alice's Entanglement object, in turn forwards the message to the stub object, which sends the message on to Bob. Note that if the second qubit in the system were still local, the process would follow the same steps without network transmission.

Now Alice's measured value of 1, along with her opHistory, is sent to Bob. Bob inverts his own opHistory, rotating by $-\pi/3$. He then looks up Alice's result in $M_e$. Since Alice measured 1, Bob will see that in the original entanglement, his only option was to get a 1. He adjusts the probabilities on his qubit accordingly. Next, Bob applies Alice's opHistory, a Sigma-X gate, to change his qubit from $|1\rangle$ to $|0\rangle$. Finally, Bob re-applies his own opHistory, rotating by $\pi/3$. This sets Bob's qubit in the correct state, $cos(\pi/3)|0\rangle + sin(\pi/3)|1\rangle$ as shown in section \ref{sec:asyncOperations}.

Indeed, we can show the equivalence:
\[
\left( \sigma_x \otimes I \right)
\left( \begin{array}{c}
1/sqrt(2)\\
0\\
0\\
1/sqrt(2) \end{array} \right)
=
\]
\[
\left( \begin{array}{cccc}
0 & 1 & 0 & 0 \\
1 & 0 & 0 & 0 \\
0 & 0 & 0 & 1 \\
0 & 0 & 1 & 0 \end{array} \right)
\left( \begin{array}{c}
1/sqrt(2)\\
0\\
0\\
1/sqrt(2) \end{array} \right)
=
\]
\[\left( \begin{array}{c}
0\\
1/sqrt(2)\\
1/sqrt(2)\\
0 \end{array} \right) \\
\]

\section{Experimental Results}\label{sec:experimentalResultBB84}
For the experimental results, we turn to the simulator which we have created to meet the criteria of distributed network entanglement simulation. Consider a system $R$ such that qubits $P$ and $Q$ were entangled in the state $1/\sqrt{2} |00\rangle + 1/\sqrt{2} |11\rangle$. If $P$ is measured as 0 in the computational basis, $Q$ will be measured as 0. The inverse is also true. Now consider a rotation of 60 degrees is applied to qubit $P$; if $P$ is measured as 0, $Q$ will have a $3/4$ chance of being measured as 0, but also a $1/4$ chance of being measured as 1. And the inverse is true here as well \cite{Veritasium}.

The results of our simulations, using the method described in section \ref{sec:quantumEntanglement}, are shown in the table below. The experiment applied a rotation of $\theta$ to $P$ and measured $P$ in the computational basis and displayed the probability amplitudes and probabilities of collapses (square of the amplitudes) for $Q$. Measurements of $P$ not equal to zero were discarded. The table below shows the predicted probability and the actual observed probability.

\begin{tabular}{l c c c c}
\hline
$\theta_P$ & sim $Q_0$ & actual $Q_0$ & sim $Q_1$ & actual $Q_1$\\
\hline
0 & 1 & 1 & 0 & 0\\
$\pi/12$ & .933 & .933 & .067 & .067\\
$\pi/6$ & .75 & .75 & .25 & .25\\
$\pi/4$ & .5 & .5 & .5 & .5\\
$\pi/3$ & .25 & .25 & .75 & .75\\
$5\pi / 12$ & .067 & .067 & .933 & .933\\
$\pi/2$ & 0 & 0 & 1 & 1\\
\end{tabular}
\captionsetup{font={small,it},singlelinecheck=off}
\captionof{table}{Probabilities for $Q=0 (Q_0)$ and $Q=1 (Q_1)$, given rotation of $P$ by $\theta$. ``Sim" values given by the simulation method presented in this paper. ``Actual" values correspond to linear expansion method.}

\section{Simulating Amplitude Damping}\label{sec:simulatingAmplitudeDamping}

Since QooSim is written in C++, it models quantum systems through the use of vector qubits and matrix algebra. These mechanics are abstracted from the developer's view by higher level objects, such as qubits and gates.

Developers create high level algorithms, such as BB84, by implementing the iRunnable interface with their algorithm class. A central singleton System object is responsible for executing the runnable algorithm class as well as managing quantum communication channels. Developers may configure some properties of these quantum configuration channels, such as signal loss or distortion and the presence of an eavesdropper.

For the simulations associated with this paper, we created two runnable classes, one called bb84Generator\_runnable, which represents Alice, and another called bb84Determiner\_runnable, which represents Bob. These runnables make use of the built-in Hadamard and $\sigma_x$ gates to manipulate qubits. Attenuation was modeled using the built-in amplitude damping class. Amplitude damping is a major factor affecting the stability of quantum systems, associated with photonic interaction with fiber material or atoms in the air \cite{nielsenChuang}. In brief, amplitude damping has the effect of causing the qubit to be lost due to interaction with the environment.

To simulate amplitude damping on a qubit, the qubit is represented using a density matrix, $\rho$. Amplitude damping is applied to this matrix through the operator-sum method, $E(\rho) = \sum_i^n E_i \rho E_i^\dagger$, where $E_n$ is the matrix associated with the $n^{th}$ outcome occurring. $E_0$ is the operator for attenuation not occurring, and $E_1$ is the operator if it has. These operators are given in equation \ref{eq:kraus_ad} \cite{nielsenChuang} \cite{anirban_1}. If attenuation has occurred, the qubit is supposed to be lost; measurement operations will not result in a zero, but will have no defined outcome.

\begin{equation}
\label{eq:kraus_ad}
E_{AD_0} =
\left( \begin{array}{cc}
1 & 0 \\
0 & \sqrt{1-\eta} \end{array} \right) ;
E_{AD_1} =
\left( \begin{array}{cc}
0 & \sqrt{\eta}) \\
0 & 0 \end{array} \right)
\end{equation}

When an eavesdropper is present, the simulator simply performs a silent measurement of the qubits passing through the quantum channel. These measurements take place at a user-configurable rate. Through this configuration, it is possible to simulate an aggressive eavesdropper, attempting to gain as much information as possible, or a stealthier eavesdropper, whose goal is to gain some information while remaining undetected. If the eavesdropper measures the qubit to be one, she transmits a new qubit set to one. Otherwise, she retransmits a qubit set to zero.

\section{Test Cases for Amplitude Damping}
\label{sec:analysis}

For our analysis, we used the simulator to run three different series of tests. In the first series, we created a baseline for analysis by simulating the effect of amplitude damping on communication through the quantum channel without an eavesdropper. In the second series, the eavesdropper was simulated close to the sending (Alice) end of the channel by applying the eavesdropper procedure followed by amplitude damping. In the last series, we added an eavesdropper close to the receiving (Bob) end of the channel. In this case, the amplitude damping procedure was applied before eavesdropping.

\begin{figure*}[t]
\begin{center}
\includegraphics[width=\linewidth]{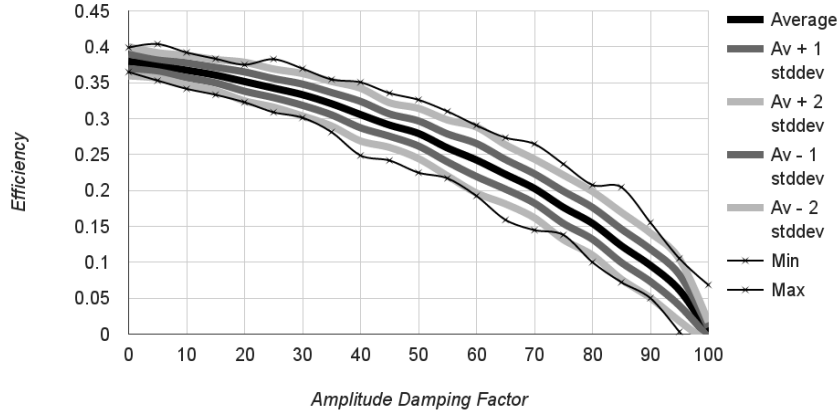}
\end{center}
\captionsetup{font={small,it},singlelinecheck=off}
\caption[BB84 efficiency vs an amplitude damping (best fit).]{
BB84 Efficiency against an amplitude damping, lines of fit for average, minimum and maximum values at each interval.
}
\label{fig:bb84_ad_bestfit}
\end{figure*}

In all three series, the amplitude damping factor represents the percentage of qubits affected by amplitude damping. This factor was varied from zero to one hundred in increments of five. At each increment, one hundred test cases were executed, for a total of two thousand cases per ``scenario." In the control series, we have only the one scenario, with no eavesdropper. In each of the series where an eavesdropper was simulated, the eavesdropper's aggressiveness was varied from ten to fifty percent of qubits measured, in increments of ten percent. We additionally simulated the one hundred percent eavesdropping scenario, because of the interesting results for series three. Each of these increments was simulated per the same ``scenario" protocol, giving twelve-thousand data points for each of the last two series.

For the control series, represented in figure \ref{fig:bb84_ad_bestfit}, we mapped the line of best fit for average performance, one and two standard deviations, and boundaries for minimum and maximum values. The outcome is now a gently sloping parabolic curve, terminating in zero efficiency at an amplitude damping factor of one hundred percent. It's useful to understand from this scenarios the correlation between distance, noise, and efficiency. The test factors ~0.68, .9, and .99 correspond to noise levels of 5, 10, and 20 dB, based on equation \ref{eq:db}. Amplitude damping measured in decibels increases linearly with distance, which implies that the effects on qubits within the channel increase exponentially. Thus, these values correspond to distances of $d$, $2d$, and $4d$, where $d$ is dependent on the noise rate on the channel.

\begin{equation}
\label{eq:db}
dB = 10 \times log_{10}(bits\ recieved / bits\ sent)
\end{equation}

\begin{figure*}[t]
\begin{center}
\includegraphics[width=\linewidth]{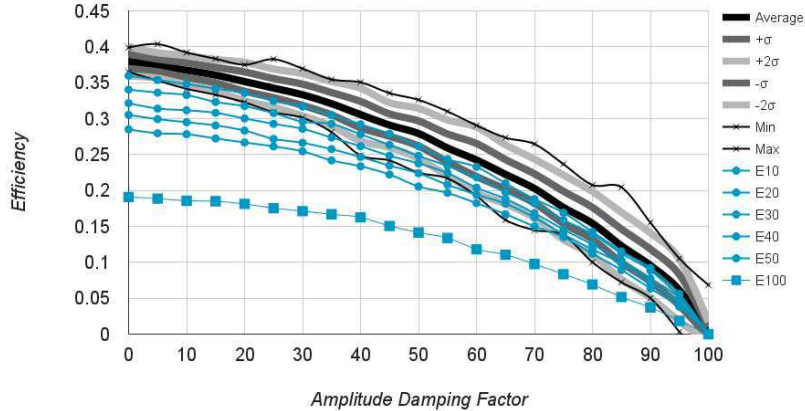}
\end{center}
\captionsetup{font={small,it},singlelinecheck=off}
\caption[BB84 Efficiency against amplitude damping compared to eavesdropping.]{
BB84 Efficiency against amplitude damping with eavesdropping on the sender's (Alice's) end. Amplitude damping varies from 0 to 100 along the horizontal axis, while eavesdropping is held constant. The lines labeled as En mark eavesdropping at a rate of n.
}
\label{fig:bb84_ad_vs_eve_alice}
\end{figure*}

In the second series, the eavesdropper was simulated near Alice's end of the communication channel. Figure \ref{fig:bb84_ad_vs_eve_alice} shows the results of this series superimposed over the control series. Eavesdropper scenarios of ten, twenty, thirty, forty, and fifty percent measurement rates are shown using thin lines accented by circles. Finally, the scenario in which the eavesdropper intercepts and measures every qubit is represented by a thin line accented with squares.

In this series, we can demonstrate that in an ideal, damping free channel, the errors introduced by an eavesdropper will likely be detectable as lower than expected channel efficiency. However, as more noise is introduced on the channel, the eavesdropper curves conform more and more neatly to the control case. At an amplitude damping factor of 10, which corresponds to .45 dB or roughly 3 km over standard fiber \cite{fiber_250}, the mean of the 10\% eavesdropping scenario is within two standard deviations of the average control case. At an amplitude damping factor of 95 (13 dB or just under 80 km), even the 100\% eavesdropping scenario is within this bound. Therefore, caution should be used if BB84 is to be implemented on links longer than this distance using current fiber technology. Shorter links may also be vulnerable if they are sufficiently noisy. 13 dB may be considered an unsafe noise level at which to operate.

\begin{figure*}[th!]
\begin{center}
\includegraphics[width=\linewidth]{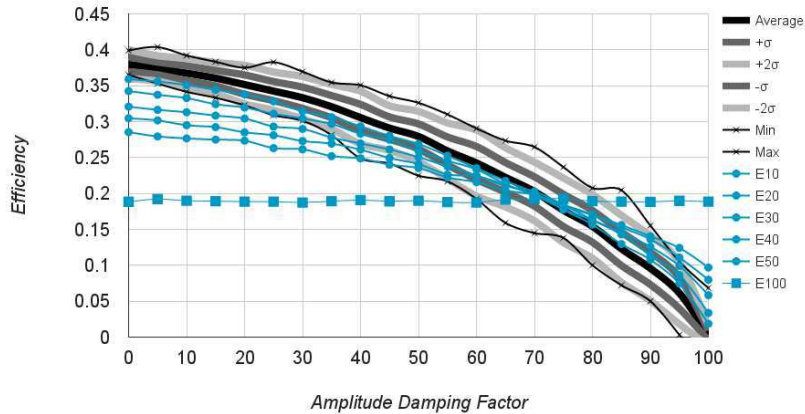}
\end{center}
\captionsetup{font={small,it},singlelinecheck=off}
\caption[BB84 Efficiency against amplitude damping compared to eavesdropping.]{
BB84 Efficiency against amplitude damping with eavesdropping on the recipient's (Bob's) end. Amplitude damping varies from 0 to 100 along the horizontal axis, while eavesdropping is held constant. The lines labeled as En mark eavesdropping at a rate of n.
}
\label{fig:bb84_ad_vs_eve_bob}
\end{figure*}

In the final series, we placed the eavesdropper near Bob's end of the communication channel. Again, in figure \ref{fig:bb84_ad_vs_eve_bob} the eavesdropper scenarios are superimposed on the control series as in the second series. Here, we see that as the rate of eavesdropping increases, it has a surprisingly positive effect on the overall efficiency of the protocol, in terms of reduced error rate. The reason for this behavior is that the eavesdropper's measurement and re-transmission of the qubit serves to ``refresh" it into a quantum state. In fact, we see that at an eavesdropping rate of one hundred percent, the protocol maintains a twenty percent efficiency. Typically, we think of eavesdropper detection in BB84 as monitoring for decreased protocol efficiency. However, in the case of a channel heavily affected by amplitude damping where the eavesdropper is close to the recipient, it may be more appropriate to monitor for unexpectedly increased efficiency.

\section{Practical Implications}
\label{sec:implications}

We now interpret our results in the context of two real world implementations of BB84. The DARPA Quantum Network \cite{boston_1, boston_2}, located in the Boston area, consists of several nodes organized in a hierarchical structure across three geographic locations; Boston University (BU), Harvard University, and BBN Technologies. The BBN site contains internal and external nodes. The internal nodes perform a sort of proxied key agreement with the BU and Harvard sites through the external nodes.

The BBN to BU link covers 19.6 km, and has an estimated attenuation of 11.5 dB. Reportedly, additional inefficiencies in the detectors at the BU recipient make quantum key distribution involving this node impossible \cite{boston_1}. Assuming these inefficiencies could be corrected, based on the link attenuation, we would model this portion of the network as having an amplitude damping factor of around 93.2. If an eavesdropper is present near the recipient (BU) end of the key exchange, the channel is heavily enough affected by amplitude damping that we would expect her to be indicated by a better than expected efficiency on the link. If the eavesdropper is instead near the sending (BBN) end of the key exchange, it appears that the error rate experienced will be near two standard deviations from average. An eavesdropper could intercept as much as fifty percent of the message without causing errors greater than one standard deviation from average. It would be extremely difficult to distinguish an eavesdropper from normal channel noise on this link.

The Harvard to BBN link is just over half the length of the BBN to BU link, at 10.2 km. Attenuation on this link is estimated at 5.1 dB, which we model with an amplitude damping factor of 69.1. Under these conditions, the errors introduced by an eavesdropper on the recipient (BBN) end of the link fall within one standard deviation from average. It would again be very difficult to detect an eavesdropper in this scenario. On the other hand, a sufficiently aggressive eavesdropper near the sending (Harvard) end of the link should be detectable. An eavesdropper who measures fifty percent or more of the qubits leaving the Harvard facility will cause errors greater than two standard deviations from average, and should cause the parties to discard their messages as compromised.

The second implementation represents one of the longest distance quantum key distribution experiments to date. Using ``ultra low loss" fibers and specialized detectors, Stucki et. al. were able to perform key exchange at distances of 100km and 250km \cite{fiber_250}. The 250 km link was estimated to have a noise ratio of 42.6 dB, which we model as an amplitude damping factor of 99.4. Under these conditions, it appears that it would be difficult to detect an eavesdropper on the sender's side of the link, even if that eavesdropper intercepts one hundred percent of the message. The errors caused by the eavesdropper will be within one standard deviation of the average for the control case. However, if the eavesdropper is present on the recipient's end of the link, she will cause the channel to be greater than one standard deviation more efficient than expected at an eavesdropping rate of twenty percent or more.

\section{Conclusion}
\label{sec:conclusion}

In this paper, we introduced the challenge of simulating quantum entanglement across a distributed network. We briefly discussed the usefulness of such a simulator in building or researching entanglement based information theory.

Conventionally, entangled qubits are manipulated through operations on the entangled state as a whole. In a distributed simulator, it is difficult and inefficient to maintain state across a network, within which either endpoint could operate on one of the qubits in the entanglement. Thus, it is desirable to develop a method for performing these operations asynchronously. We showed a method for performing quantum gate operations in isolation with reconciliation on measurement. We demonstrated through direct proof that this method is equivalent to the conventional method, and provided experimental data to substantiate this claim. By using our method for asynchronous entanglement operations, we have created a toolkit for learning and demonstration of quantum protocols and concepts.

In the latter half of the paper, we have demonstrated, through simulation, the combined effects of channel attenuation through amplitude damping and eavesdropping on the efficiency and security of the BB84 quantum key distribution protocol. An interesting aside to this research was the realization that the position of the eavesdropper may influence the overall efficiency of the channel, as she may act as a kind of repeater. Consequently, parties participating in BB84 key exchanges should be wary of efficiencies both below and above expectations.

We have an open question as to what method to use to establish the expected channel noise rate. For example, a clever attacker could cause higher than normal attenuation on the link in order to afford herself more opportunities to eavesdrop undetected. It is often noted that the parties must pre-establish an acceptable error rate, but we have yet to find a description of a procedure to do this.

However this question is answered, it is clear to us that there are noise thresholds above which the BB84 protocol should not be used on a channel. Since noise is highly influenced by the length of the channel, noise limits the ``safe" range at which the protocol may be used. It is likely that mechanisms to eliminate or compensate for amplitude damping will be needed for this range to be extended.


\end{document}